\documentclass[prl,aps,superscriptaddress,showpacs,amsmath,twocolumn]{revtex4}

\usepackage[dvips]{graphicx}
\usepackage{dcolumn}
\usepackage{bm}
\usepackage{amsmath}

\begin{document}




\title{ Stable liquid Hydrogen at high pressure by   
a novel  \emph{ab-initio} molecular dynamics}  

\author{ Claudio Attaccalite}
\affiliation{Institute for Electronics, Microelectronics, and Nanotechnology Dept.  ISEN B.P. 60069 59652 Villeneuve d'Ascq Cedex France} 
\author{Sandro Sorella}
\affiliation{Democritos National Simulation Center, 
SISSA, Via Beirut n.2, Trieste, Italy} 
\date{\today}
\begin{abstract}
We introduce an efficient scheme for the molecular dynamics 
 of electronic systems by means of  quantum Monte Carlo. 
The evaluation of 
the (Born-Oppenheimer) forces acting on the ionic positions 
is achieved by two main ingredients: i) 
the forces are computed with finite and small variance,  which allows  
the simulation of a  large number of atoms, 
ii) the statistical noise 
corresponding to the forces is used to drive the dynamics 
at finite temperature by means of  an appropriate Langevin dynamics. 
A first application to the high-density phase of Hydrogen is 
given, supporting the stability of the  
 liquid phase  at $\simeq 300GPa$ and $\simeq 400K$. 
\end{abstract}

\maketitle

\narrowtext
The phase diagram of Hydrogen at high pressure is still under intense 
study from the experimental and theoretical point of view.
In particular in the low temperature high-pressure regime there is yet no clear evidence of a metallic atomic solid, 
and either a molecular solid phase can be favoured even 
in this regime\cite{needs}, or
 the liquid phase can be stabilized at low temperature  by
 increasing  the  pressure (see Ref.(\cite{galli})).

Indeed, for  high pressures around $300GPa$,  
a two fluid (proton and electron) superconducting  phase,  
induced by the  strong electron-phonon coupling, 
has been conjectured\cite{ash1}  and 
unusual quantum properties have been later predicted\cite{ashcroft}.
In this work we use an improved  {\em ab}-initio molecular dynamics (AMD) 
by using accurate forces computed by Quantum Monte Carlo (QMC).   
We present preliminary 
results, showing  that the   liquid phase is energetically stable, due to the strong electron correlation, at least within the Resonating Valence Bond 
(RVB)  variational approach\cite{vanilla}, 
which is very accurate also in the solid phase.

AMD is well established as a powerful tool to investigate many-body condensed matter systems. 
Indeed, previous attempts to apply  Quantum Monte Carlo (QMC) 
 for   the dynamics of ions\cite{mitas} 
or for their thermodynamic properties\cite{penalty} are known, but they 
were limited to small number  $N$ of electrons  or to 
total energy corrections  
of the  AMD trajectories, namely without the explicit 
calculation of the forces. Indeed,  the technical achievements that 
we are going to present in this letter 
are  particularly important for the simulation of 
liquid or disordered phases by QMC.

\vskip 2mm \noindent {\it Calculation of forces with finite variance.}\hskip 2mm
The simplest method  for accurate calculations within QMC, is given by 
the so called variational Monte Carlo (VMC), which allows to compute the 
variational energy expectation value $E_{VMC}={ \langle \psi_T |H | \psi_T \rangle \over \langle \psi_T | \psi_T \rangle }  $ 
of a highly accurate correlated wave function (WF)  $\psi_T$ by means of 
a statistical approach:
electronic configurations $\left\{ x \right\}$, with given 
electron positions $\vec r_i$  and spins $\sigma_i=\pm 1/2$ for $i=1,\cdots N$, 
 are usually generated 
by the Metropolis algorithm according to the probability density 
$\mu_x \propto \psi_T(x)^2$. Then $E_{VMC}$ is computed by averaging 
statistically over $\mu_x$ 
the so called local energy $e_L(x)= { \langle  \psi_T |H 
| x \rangle  \over \langle \psi_T | x \rangle } $, namely 
 $E_{VMC} = \int d\mu_x~ e_L(x)$, 
where $\int d \mu_x $ indicates conventionally the $3 N$ multidimensional 
integral over the electronic coordinates weighted by $\psi_T^2(x)$.
In the present work we assume that the WF  $\psi_T(x)= \langle x |
\psi_T \rangle=  J  \times 
 \det A$ is given by a correlated Jastrow factor $J$ 
times a determinant $D$ of a $N \times N$ matrix $A$, such 
as for instance a Slater determinant. 
The main ideas  of this approach can be straightforwardly generalized to more 
complicated  and more accurate WF's, as well as to  
projection QMC 
 methods\cite{reptation}  more accurate than VMC. 

The efficient calculation of the energy derivatives, namely  the forces 
 $ \vec f_{\vec R_i}  =- { \partial E_{VMC}  \over  \partial \vec R_i }$, 
for $i=1,\cdots N_A$, where $N_A$ is the number of atoms, 
is the most important ingredient for the AMD.
Within VMC they can be computed by simple differentiation 
of $E_{VMC}$, using that not only the Hamiltonian $H$ but also  
$\psi_T$ depend explicitly on the  
atomic positions $\vec R_i$. This leads to two different  contributions 
to the 
force $\vec f_{\vec R_i} = \vec f^{HF}_{\vec R_i} + \vec f^P_{\vec R_i}$,  
the Hellmann-Feynman  $\vec f^{HF}$ and the Pulay one 
$\vec f^P_{\vec R_i}$,  where:
\begin{eqnarray}
\vec f_{\vec R_i}^{HF} & = & -\int d\mu_x ~    \langle x | 
\partial_{\vec R_i} H | x \rangle  \label{HF}   \\
\vec f_{\vec R_i}^{P} & = & -2 \int d\mu_x ~ (e_L(x) -E_{VMC})   
\partial_{\vec R_i} {\rm log}|\psi_T (x)|   \label{P}  
\end{eqnarray}
However in order to obtain a  statistically meaningful average, namely with 
finite variance, 
some manipulations  are  necessary because the first integrand  
may diverge when the minimum electron-atom distance vanishes,
 whereas the second integrand is analogously
unbounded  when an electronic  configuration $x$
approaches the nodal surface   determined by $\psi_T(x)=0$.
By defining   with  $d$ ($\delta$) 
 the distance of $x$ from the nodal region (the minimum 
electron-atom distance), 
 $e_L(x),  \partial_{\vec R_i} {\rm log}  \psi_T (x)  \simeq 1/d$ 
( $\langle x | \partial_{\vec R_i}  H | x \rangle  \simeq 1/\delta^2$), whereas 
$\mu_x \simeq d^2$ ($\mu_x \simeq \delta^2$), 
 leading to an unbounded integral of the 
 square integrand in Eq.(\ref{P}) (Eq.\ref{HF}),
 namely to {\em infinite variance}.  
The infinite variance problem in Eq.(\ref{HF}) was solved in several 
ways. Here we adopt a very elegant and efficient scheme 
proposed by Caffarel and Assaraf\cite{caffarel}. 
Instead the infinite variance problem in 
Eq.(\ref{P}) was not considered so  far, 
 and this is clearly a problem for  a meaningful 
definition of ionic   AMD consistent with QMC forces.

In this letter we solve this problem in the 
following simple way, by  using the so called re-weighting method.
We use a different probability distribution 
$\mu^\epsilon_x \propto \psi_G(x)^2$, determined by a guiding 
function $\psi_G(x)$:
\begin{equation}
\psi_G(x) = R^\epsilon (x) ( \psi_T(x)/R(x)) 
\end{equation}
where $R (x) \propto \psi_T(x) \to 0$  for $d \to 0$ 
is a ''measure'' of the distance from  the nodal surface $\psi_T(x)=0$.
By assumption $\psi_T$ may vanish  only when  $\det A=0$  
($J>0$) and therefore $R(x)$ is  chosen to depend only on $A$.  
 For reasons that will become clear later on we have adopted  the following 
expression:
\begin{equation}
R(x) = 1/\sqrt{ \sum\limits_{i,j=1}^N  |A^{-1}_{i,j}|^2}. \label{defr}
\end{equation}
Then the guiding function is defined  by properly regularizing $R(x)$, 
namely:
\begin{equation} \label{reg}
R^\epsilon(x) = \left\{ 
\begin{array}{cc}
R(x) & {\rm if ~}  R(x) \ge \epsilon \\
\epsilon (R(x)/\epsilon)^{ R(x)/\epsilon}  & {\rm if ~}  R(x) < \epsilon \\
\end{array} 
\right..
\end{equation}
The non obvious regularization for $ R(x) < \epsilon$ 
instead of e.g. 
$R^\epsilon (x)=Max[\epsilon, R(x)]$ was considered in order to 
satisfy the continuity of the first derivative of $\psi_G(x)$ when 
$R(x)=\epsilon$, thus ensuring that $\psi_G(x)$ remains as close as 
possible to the trial function $\psi_T$.   
In this way the Metropolis algorithm can be applied for generating 
configurations  according to a slightly different 
 probability  $\mu^\epsilon(x)$ 
and the exact expression 
of $\vec f^{P}_{\vec R_i}$ can be obtained by the so called umbrella 
average:
\begin{equation} \label{umbrella}
\vec f_{\vec R_i}^{P}  = {  -2 \int d\mu^\epsilon_x ~ S(x) (e_L(x) -E_{VMC})   
\partial_{\vec R_i} log \psi_T (x)   
\over \int d\mu^\epsilon_x ~ S(x) }.
\end{equation}
Now,  
 the re-weighting factor $S(x)= (\psi_T(x)/\psi_G(x))^2=
 Min \left[ 1, (R(x)/\epsilon)\right]^{2 -2 R(x)/\epsilon}  \propto d^2 $, 
cancels out the divergence of the integrand, that was instead present in 
Eq.(\ref{P}).
Hence the mentioned integrands  in the numerator 
and $S(x)$ ( $  \le 1$)  in the denominator 
of Eq.(\ref{umbrella}) represent  bounded random variables and 
have obviously {\em finite variance}.
In this way  the problem of infinite variance is definitely 
solved within this simple re-weighting scheme. 

\begin{figure}[hbt]
\includegraphics[width=0.5\textwidth,angle=0]{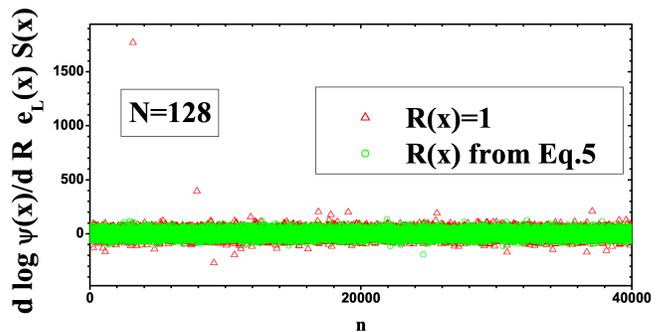}
\begin{center}
\caption{(color online). 
Evolution of the integrand in Eq.(\ref{umbrella}) as a function 
of the Monte Carlo iterations. Each new sample  is obtained after  
$2 N$ Metropolis trials.}
\label{fig1}
\end{center}
\end{figure}

We show in Fig.(\ref{fig1}) the efficiency of the  method for  
computing the Pulay force component acting on a   
Hydrogen proton at $r_s=1.31$ in a bcc lattice. 
As it is clear in the plot for the $N=128$ case, 
the difference between a method with finite 
variance 
and the standard one with infinite variance is evident. 
In this way 
 the $3 N_A \times 3 N_A$ correlation  matrix 
$\bar \alpha_{QMC}$,  defining  the statistical correlation 
between the force components, can be efficiently evaluated: 
\begin{equation}\label{noiseqmc}
\bar  \alpha_{QMC} (\vec R) =  <   (\vec f_{\vec R_i} - < \vec f_{\vec R_i} >)  
( \vec  f_{\vec R_j} -< \vec f_{\vec R_j}> )  >  
\end{equation}
where   the brackets $< >$ 
 indicate the statistical average over the QMC samples.
The correlation matrix  $\bar \alpha_{QMC}$, that  
within the conventional  method is not  even defined, 
will be a fundamental ingredient 
for a consistent AMD  with QMC forces and therefore the 
solution of the infinite variance problem 
 is crucial for this purpose.

 \vskip 2mm \noindent {\it  Langevin dynamics.} \hskip 2mm
In the following derivation we  assume that ions have 
unit mass, that can be generally  obtained by e.g. 
a simple rescaling of lengths for each ion independently. 
 For clarity and compactness of notations,
we also  omit the ionic subindices $i$  when  
not explicitly necessary.
Moreover   matrices (vectors) 
are indicated by a bar (arrow)  over the corresponding 
symbols, and  the matrix-vector product is also implicitly understood. 
We start therefore by the   following 
  AMD  equations for the 
ion coordinates $\vec R$ and velocities $\vec v$:
\begin{eqnarray} \label{ldyn}
 \dot {\vec v }  &=&  -  \bar \gamma(\vec R) 
  \vec v   +   \vec f(\vec R)   +\vec \eta  (t)   \\
 \dot{  \vec R }   &=&  \vec v    \label{dynR} 
\end{eqnarray} 
By using the fluctuation-dissipation theorem 
the friction matrix $\bar \gamma$ is related to the temperature 
$T$ (henceforth the Boltzmann constant $k_B=1$) by:
\begin{equation} \label{eqpar}
\bar  \gamma(\vec R) = { 1 \over 2 T } \bar  \alpha (\vec R)
\end{equation}
where  $ \bar \alpha (\vec R)$ is generally a 
 symmetric correlation matrix:
\begin{equation}
<\vec \eta_i(t) \vec \eta_j(t^\prime)>=\delta(t-t^\prime) \bar \alpha (\vec R). 
\end{equation}
It is important to emphasize that, as a remarkable generalization 
of the standard AMD  used in \cite{parrforce}, 
in the present approach  the 
friction matrix $\bar \gamma$, 
may depend explicitly on the ion positions $\vec R$, 
so that Eq.(\ref{eqpar}) can be satisfied 
even for a generic correlation matrix $\bar \alpha (\vec R)$. 
In fact we have the freedom to consider  
a QMC contribution in  $\bar \alpha (\vec R)$:
\begin{equation} \label{forces}
\bar \alpha (\vec R)  = \bar \alpha_{0} + \Delta_0 \, 
 \bar \alpha_{QMC} (\vec R)  
\end{equation}
where $\Delta_0>0$ and 
$\bar \alpha_{0}$ is the identity 
matrix $\bar I$  up to another  positive constant $\alpha_0$,
$\bar \alpha_{0} = \alpha_0 \bar I $, and $\alpha_{QMC}(R)$ 
can be estimated by Eq.(\ref{noiseqmc}). 
In the following we will show that, for appropriate $\alpha_0,\Delta_0>0$, 
it is possible to follow the Langevin dynamics  
by means of noisy QMC forces.

\vskip 2mm \noindent {\it Integration of the Langevin dynamics.} \hskip 2mm
Henceforth  the velocities 
$v_n$ are computed at half-integer  times $t_n-{ \Delta t \over 2}$, whereas 
coordinates $\vec R_n$ are assumed to be defined at integer times 
$\vec R_n= \vec R(t_n)$. 
Then,  in the interval $ t_n-{ \Delta t \over 2} 
  < t < t_n+{ \Delta t \over 2}$ and  for $\Delta t$ small, 
the positions 
$\vec R$ are changing a little and, within a good approximation, 
 the $\vec R$ 
dependence in the  Eq.(\ref{ldyn}) can be neglected, so that this differential 
equation becomes  linear and can be solved explicitly.
The closed solution 
 can be recasted  in the following useful  form, 
where the force components appear 
  corrected by appropriate noisy vectors $\vec {\tilde \eta}$:
\begin{eqnarray} \label{MD}
\vec v_{n+1}  &=&e^{ - \bar \gamma \Delta t }  \vec v_{n} +
\bar \Gamma  ( \vec f(\vec R_n) +   \vec { \tilde \eta} )   \\
\vec R_{n+1} & =& \vec R_n+  \Delta t \,\vec v_{n+1} +O(\Delta t^3) \label{eqappr}\\
\bar \Gamma &=& \bar \gamma^{-1} ( 1 - e^{ -\bar \gamma \Delta t } )  \\ 
{\vec {\tilde \eta} } &=& 
{ \bar \gamma  \over  2  \sinh( {  \Delta t \over 2}  \bar \gamma) } 
\int\limits_{t_n- {\Delta t \over 2} }^{t_n+{ \Delta t \over 2} } dt e^{ \bar \gamma (t-t_n)} \vec \eta (t)    \label{totnoise} 
\end{eqnarray}
By using  that
$\bar \alpha = 2 T  \bar \gamma$ from Eq.(\ref{eqpar}), 
and that its dependence on  $\vec R$ can be consistently neglected 
in this small time interval,  
 the correlator defining the discrete (time integrated) 
noise  $ \tilde {\vec \eta} $ can be computed explicitly:
\begin{equation} \label{totalnoise}
<  \vec {\tilde \eta}_i \vec {\tilde \eta}_j > = 
{ 2 T } \bar \gamma^2  {  \sinh ( \Delta t \bar \gamma  
) \over
 4 \sinh(  { \Delta t  \over 2} \bar \gamma)^2 }= \bar \alpha^{\prime} 
\end{equation} 
This means that 
 the QMC noise has to be corrected in a non trivial way as explained in the 
following.   

\begin{figure}[hbt]
\includegraphics[width=8.cm,angle=0]{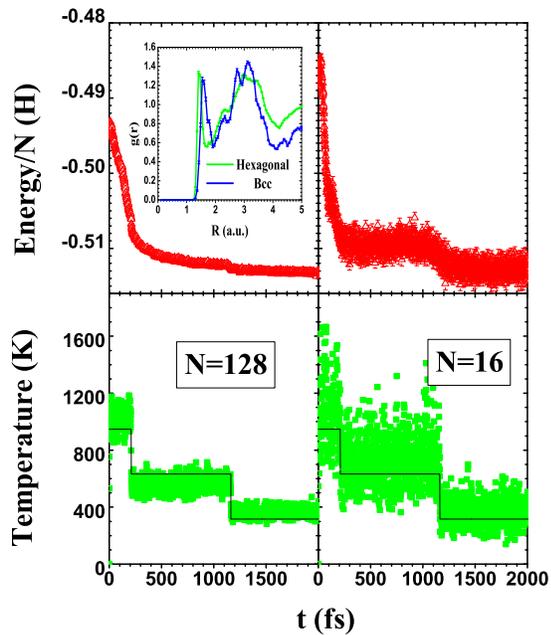}
\begin{center}
\caption{(color online). Evolution of the internal energy and temperature 
vs the AMD with QMC forces.
$\Delta t= \Delta_0=1.036fs$, 
$\alpha_0=0.7 k_B T a.u.$. Bottom: points represent instantaneous temperatures 
estimated by the average kinetic energy, lines represent the target 
 temperatures. They should coincide on average for $\Delta t \to 0$. 
 The average energy, pressure and temperature in the last 0.5ps are
$-0.51319 \pm 0.00003 H$  ($-0.5127\pm 0.0001 H$)  
$364K \pm 5K$  ($364 \pm 10K$)   and 
$335\pm 2 GPa$ ($394\pm5 GPa$)  for $N=128$ ($N=16$), 
respectively. At each iteration the statistical noise on the total 
energy is $\simeq 5000K$. The proton-proton   
  $g(r)$, averaged at the lowest temperature,  is shown in the inset.}
\label{fig2}
\end{center}
\end{figure}

\vskip 2mm  \noindent {\it Noise correction.} \hskip 2mm
The QMC noise is 
given during the simulation, and therefore in order to follow the 
correct dynamics another noise $\vec \eta^{\,ext}$ has to be added
to the noisy force components in a way  that the total integrated noise 
is the correct expression (\ref{totalnoise}), i.e.  
$ { \vec {\tilde  \eta} } = { \vec \eta}^{\,ext}+ \vec \eta_{QMC}.$
By using that the QMC noise 
in  Eq.(\ref{noiseqmc}) is obviously 
independent of the external noise, we easily obtain  
 the corresponding correlation matrix:
\begin{equation} \label{extnoise}
< \vec  \eta^{\,ext}_i \vec \eta^{\,ext}_j > = 
  \bar \alpha^{\prime} -\bar \alpha_{QMC} 
\end{equation}
On the other hand, after substituting the expression
  (\ref{forces}) in Eq.(\ref{eqpar}) 
$\bar \gamma={ 1\over 2 T }
 (\bar \alpha_0 + \Delta_0 \,\bar \alpha_{QMC})$
and using   the expression  (\ref{totalnoise}) for  
$\bar \alpha^\prime$, we obtain  a positive definite matrix 
in Eq.(\ref{extnoise}) for $\Delta t\le \Delta_0$ \cite{proof}. 
Hence $\vec \eta^{\,ext}$ is  
a generic Gaussian correlated noise 
that can be easily  sampled  by   standard algorithms. After that  
 the   random vector $\vec \eta^{\,ext}$     is   added    
 to the force $\vec f + \vec  \eta_{QMC}$ 
obtained by QMC,
and  replaces $\vec f + \vec { \tilde  \eta}  $  in Eq.(\ref{MD}). 
 This  finally allows to obtain 
an accurate AMD  with a corresponding 
small time step error. 
The main advantage of this technique is that, at each iteration, 
by means of 
Eq.(\ref{eqpar}),
the statistical noise on the total energy 
 and forces (see Fig.\ref{fig2})
  can be much larger than the target temperature
 $T$, 
and this allows to improve dramatically the QMC efficiency. 
\vskip 2mm \noindent {\it Optimization of  the WF.}  \hskip 2mm
In the following examples we consider a cubic box   
with periodic boundary conditions, and use 
a variational WF $J \times \det A$ 
  that 
is able to provide a very accurate description of the
correlation energy, due to 
a particularly efficient  
choice of the determinant factor, that allows to describe the RVB
 correlations \cite{casulamol,benzenedim}.
The WF contains  several variational parameters, indicated by a 
vector  $\vec \beta $, 
 that have to be consistently optimized during the AMD. 
The Jastrow factor $J$ used here  depends  both on the charge and spin
densities, and it is expanded in a localized atomic basis.
As it is shown in the table, the accuracy of our WF is 
remarkable.
Indeed the small difference between the so called DMC
-providing the lowest possible  variational energy within
the same nodal surface of $\psi_T$-
 and the VMC energies 
clearly supports the accuracy of our calculation. 
The size effects are very large for the metal and have been estimated 
($N=\infty$) following Ref.\onlinecite{oldh}. 
\begin{table} \caption{\label{htab} 
Comparison of the total energy per proton (Hartree) 
 for Hydrogen  in the bcc lattice at $r_s=1.31$ compared with the 
  published ones with lowest energy (to our knowledge).}  
\begin{center}
\begin{tabular}{|l|c|c|c|r|}
\hline
 N & $E_{VMC}/N_A$ & $E_{VMC}/N_A$\cite{pier} & 
 $E_{DMC}/N_A $ &  $E_{DMC}/N_A $\cite{pier}  \\
\hline 
 16  & -0.48875(5) &  -0.4878(1) &  -0.49164(4)  & -0.4905(1) \\
 54  & -0.53573(2) &  -0.5353(2) & -0.53805(4)  & -0.5390(5) \\
 128 & -0.49495(1) & -0.4947(2) &  -0.49661(3)  & -0.4978(4)  \\
 250 & -0.49740(2)  & -  & -0.49923(2) & - \\ 
 432 & -0.49943(3)  & - &  - & - \\
 $\infty$ & -0.501(1) & -  & -0.503(1) & - \\
\hline
\hline
\end{tabular}
\end{center}
\end{table}

In order to optimize the WF we use the recent 
method introduced in Ref.\onlinecite{hesscyrus}, devised here 
in an appropriate way to optimize a large number 
of parameters during the AMD simulation, as described in 
Ref.\onlinecite{benzenedim}. 
This allows to remain efficiently within the Born-Oppenheimer energy surface 
each time the ionic positions are changed  according to Eq.(\ref{MD}).

\vskip 2mm \noindent {\it Application to high-pressure Hydrogen.} \hskip 2mm
We show in Fig.(\ref{fig2}) the evolution of the internal energy
and corresponding temperature
as a function of time with the proposed AMD with  QMC forces, starting
from the bcc Hydrogen solid at $r_s=1.31$, considered henceforth.
Although we have not studied the possible stability of all other
solid phases yet,  for  $N=128$ also 
the simple hexagonal structure (SH) melts.
This already provides  a clear support to the liquid phase because
these two atomic solids are the  most stable ones, so far proposed at
zero temperature.  
 The proton-proton
correlation function obtained starting from the two atomic solids
is shown in the inset.

We performed a finite size scaling analysis based on the comparison of 
our QMC results with the 
LDA ones\cite{oldh}. We found that LDA favours atomic 
solid phases respect to QMC.
For instance the internal 
energy difference between the liquid phase and the BCC solid
one is only -0.0004 Ha in LDA while in QMC is -0.011 Ha. 
On the other hand the molecular solid (mhpc-c\cite{oldh}) is also similarly 
preferred to the atomic solid by QMC, but appears to 
have much larger  zero point energy corrections 
compared to the liquid.\cite{russo} 
Indeed we have 
 studied quantum effects on protons by using the Wigner-Kirkwood
expansion  on the free energy and obtained 
$8.5(2)mH/proton$ for the liquid, namely a   much smaller correction 
than the SH  [$12.3(2)mH/{\rm proton}$] and the 
molecular solid  [$13.1(2) mH/{\rm proton}$] ones. 
Finally  even  the more accurate DMC does not affect  
the  liquid stability, because it lowers 
 the internal energy of all the phases studied by about the same amount 
($2-3mH/{\rm proton}$).  

In conclusion we have shown that it is possible to make a realistic 
and accurate AMD simulation with QMC forces.  
We found that  
the bcc and SH solid structures appear 
clearly  unstable even at low temperatures 
where a molecular liquid 
  has  much lower internal energy, and is further stabilized by considering   
quantum effects on protons.
This important finding highlights the present QMC technique  as a 
 possible and accurate alternative to study phase diagrams of materials and as a benchmark for other approximate methods.

We acknowledge partial support by PRIN MIUR  and   CNR.
 We thank  D.M. Ceperley, C. Pierleoni, R. Car, 
F. Becca, M. Casula, M. Fabrizio 
for useful discussions, and 
the excellent stability of SP5 in CINECA.





\begin{thebibliography}{99}
\bibitem{galli} S. A. Bonev, E. Schwegler, T. Ogitsu and G. Galli, 
Nature, {\bf 431}, 669 (2004).
\bibitem{ash1} N.W. Ashcroft , J.Phys. {\bf A 12},
A129-137 (2000).
\bibitem{ashcroft}  E. Babaev, A. Sudbe, and N. W. Ashcroft,
 Nature, {\bf 431}, 666 (2004), 



\bibitem{needs} C. J. Pickard1 and R. J. Needs,
Nature Physics, {\bf 3}, 473 (2007).


\bibitem{vanilla} see e.g.  P.W. Anderson  {\it et al.} J. 
 Phys. Cond. Mat. {\bf 16} R755-R769 (2004) and references therein.

\bibitem{mitas} J. C. Grossman and L. Mitas \prl {\bf 94}, 056403 (2005)

\bibitem{penalty} C. Pierleoni, D.M. Ceperley, and M. Holzmann \prl {bf 93}, 
146402 (2004), D.M. Ceperley and M. Dewing J. Chem. Phys. {\bf 110}, 9812 (1999).

\bibitem{reptation} S. Moroni and S. Baroni \prl {\bf 82}, 4745  (1999). 

\bibitem{caffarel} R. Assaraf and M. Caffarel J. Chem. Phys. {\bf 113}, 4028 
(2000).

\bibitem{parrforce} F. R. Krajewski and Michele Parrinello, \prb 
{\bf 73}, 041105(R) (2006).


\bibitem{proof} By definition $\bar \alpha^\prime$ commutes with   
 $\bar \alpha_{QMC}$, and therefore they have common eigenvectors. On the 
other hand each eigenvalue $\lambda$ 
of $\bar \alpha_{QMC}$ is positive, and  the corresponding one for
 $\bar \alpha^\prime-\bar \alpha_{QMC}$ is  
 $ 2 T   ( { x \over 2 \Delta t  \sinh(x/2) })^2 \sinh x  -\lambda$, namely 
a monotonically increasing  function  of  
 $ x = { \Delta t  \over 2 T} (  \alpha_0 + \Delta_0 \lambda)$. 
Thus  this eigenvalue is greater than the one  
evaluated  for the   minimum $x$ value, namely for  
$\alpha_0=0$ and  $\Delta_0  =\Delta t$. 
This bound  remains  positive, vanishing   only for $\Delta t\to0$. 


\bibitem{casulamol} M. Casula, C. Attaccalite and S. Sorella J. Chem. Phys.
{\bf 121} 7110 (2004).

\bibitem{benzenedim} S. Sorella, M. Casula and D. Rocca, 
J. Chem. Phys. in press.  

\bibitem{pier} M. Holzmann, D. M. Ceperley, C. Pierleoni, and K. Esler,
\pre {\bf 68}, 046707 (2003). 

\bibitem{hesscyrus} C. Umrigar et al.,  \prl {\bf 98}, 110201 (2007).

\bibitem{oldh} V. Natoli, R. M. Martin, and D. M.  Ceperley \prl 
{\bf 70} 1952 (1993), ibidem  {\bf 74} 1601 (1995).


\bibitem{russo} V. V. Kechin, JETP Letters {\bf 79}, 40 (2004).




















\end{thebibliography}
\end{document}